\documentclass[11pt, a4paper]{article} 

\usepackage{jcappub}
\usepackage{epsf}  
\usepackage{eucal} 
\usepackage{amssymb}
\usepackage{amsmath} 
\usepackage{graphicx} 
\usepackage{bm}
\usepackage{dcolumn} 
\usepackage{epsfig} 
\usepackage{multirow}
\usepackage{hyperref}
\usepackage{color}

\begin{document}

\title{General dissipation coefficient in low-temperature warm inflation}

\author[a]{Mar Bastero-Gil}
\emailAdd{mbg@ugr.es}
\affiliation[a]{Departamento de F\'{\i}sica Te\'orica y del Cosmos, Universidad de Granada, Granada-18071, Spain}

\author[b]{Arjun Berera}
\emailAdd{ab@ph.ed.ac.uk} 
\affiliation[b]{SUPA, School of Physics and Astronomy, University of Edinburgh, Edinburgh, EH9 3JZ, United Kingdom}

\author[c]{Rudnei O. Ramos}
\emailAdd{rudnei@uerj.br}
\affiliation[c]{Departamento de F\'isica Te\'orica, Universidade do Estado do Rio de Janeiro, 20550-013 Rio de Janeiro, RJ, Brazil}

\author[b]{Jo\~ao G. Rosa}
\emailAdd{joao.rosa@ed.ac.uk} 

\date{\today}

\arxivnumber{1207.0445}

\abstract{
In generic particle physics models, the inflaton field is coupled to other bosonic and fermionic fields that acquire large masses during inflation and may decay into light degrees of freedom. This leads to dissipative effects that modify the inflationary dynamics and may generate a nearly-thermal radiation bath, such that inflation occurs in a warm rather than supercooled environment. In this work, we perform a numerical computation and obtain expressions for the associated dissipation coefficient in supersymmetric models, focusing on the regime where the radiation temperature is below the heavy mass threshold. The dissipation coefficient receives contributions from the decay of both on-shell and off-shell degrees of freedom, which are dominant for small and large couplings, respectively, taking into account the light field multiplicities. In particular, we find that the contribution from on-shell decays, although Boltzmann-suppressed, can be much larger than that of virtual modes, which is bounded by the validity of a perturbative analysis. This result opens up new possibilities for realizations of warm inflation in supersymmetric field theories.
}



\maketitle



\section{Introduction}

Inflation \cite {inflation} is an extremely successful paradigm providing an elegant solution to the shortcomings of the standard cosmological model, in particular explaining the high degree of flatness and homogeneity of the observable universe, as well as describing the origin of the temperature anisotropies observed in the Cosmic Microwave Background and the seeds of the Large Scale Structure of our universe.

In the traditional picture of inflation, the early universe is dominated by the vacuum energy of a scalar field $\phi$ that slowly rolls down its potential, $V(\phi)$. This mimics the effect of a cosmological constant if the kinetic energy of the inflaton field is negligible, a condition that is necessarily violated at some point during the evolution of the field towards the minimum of the potential, thus yielding a finite period of accelerated expansion. This typically requires a scalar field with extremely weak self-interactions, so that its very flat potential leads to 40-60 e-folds of inflation.

The quasi-exponential expansion redshifts away any matter or radiation present before the scalar field comes to dominate the energy density:
\begin{equation} \label{radiation equation}
\dot\rho_R+4H\rho_R=0 \qquad \Rightarrow\qquad \rho_R\sim e^{-4Ht}~,
\end{equation}
and consequently one needs a mechanism to reheat the cold inflationary universe and provide an exit into the standard radiation-dominated cosmology. The inflaton field must thus have some interactions with other fields, which directly or indirectly lead to the production of at least the Standard Model degrees of freedom and the particles that presumably constitute the inferred dark matter content of our universe.

These interactions must, however, be sufficiently weak in order to preserve the flatness of the inflaton potential at the quantum level, which can be achieved by either (fine-)tuning the associated couplings or by considering additional symmetries that, in particular, keep the mass of the scalar inflaton below the value of the Hubble parameter during inflation, thus ensuring an overdamped evolution. Moreover, when one considers generic couplings to additional scalar (fermionic) fields of the form $g^2\phi^2\chi^2$ ($g\phi\bar\psi\psi$), these fields typically become heavy during inflation due to the large values of the inflaton vacuum expectation value (vev). This implies that the decays $\phi\rightarrow \chi\chi, \bar\psi\psi$ are generically forbidden in the slow-roll regime, with reheating of these degrees of freedom occuring naturally only at the end of inflation if the inflaton vev drops sufficiently.

Interactions with additional scalar fields may also have interesting dynamical consequences, as is the case of hybrid inflation models \cite{Linde:1993cn, Copeland:1994vg}, where negative contributions to the squared mass of the waterfall field $\chi$, coupled to the inflaton as above, make it tachyonic at some critical value of $\phi$. Inflation then ends with a phase transition in which the $\chi$ field evolves towards the true minimum of its potential, and in the process it may reheat the universe if it is coupled to radiation.  

Warm inflation \cite{wi, Berera:1996nv} (see also \cite{earlydissp}) provides an alternative picture of the inflationary universe, in which particle production is sourced by the rolling inflaton field itself, so that radiation is not completely diluted away:
\begin{equation} \label{radiation equation_warm}
\dot\rho_R+4H\rho_R=\Upsilon\dot\phi^2 \qquad \Rightarrow\qquad \rho_R\rightarrow {\Upsilon\dot\phi^2\over 4H}\sim const.\;,
\end{equation}
in the slow-roll regime. Although accelerated expansion only occurs for a sub-dominant radiation component, $\rho_R\ll \rho_\phi$, it is possible that its temperature exceeds the de Sitter temperature, $T>H$, which significantly changes the inflationary dynamics\footnote{In this work we will consider a radiation bath that is close to thermal equilibrium, although this need not be the case in general.}. On one hand, the spectrum of primordial density fluctuations is seeded by thermal rather than vacuum fluctuations of the inflaton field \cite{wi,Berera:1999ws,Taylor:2000ze,Hall:2003zp, Moss:2008yb}, which may lead to interesting observational consequences such as the suppression of the tensor-to-scalar ratio and significant deviations from a gaussian spectrum \cite{Gupta:2002kn, Chen:2007gd, Moss:2007cv, Moss:2011qc}. On the other hand, particle production induces an additional friction term in the inflaton's motion:
\begin{equation} \label{inflaton density_warm}
\dot\rho_\phi+3H(\rho_\phi+p_\phi)=-\Upsilon\dot\phi^2~,
\end{equation}
which using $\rho_\phi=\dot\phi^2/2+V(\phi)$ and $p_\phi=\dot\phi^2/2-V(\phi)$ leads to:
\begin{equation} \label{inflaton equation_warm}
\ddot\phi+3H\dot\phi+V_\phi=-\Upsilon\dot\phi~,
\end{equation}
where $V_\phi$ denotes the derivative of the inflaton potential. This friction then contributes to overdamp the inflaton's motion, alleviating the need for very finely-tuned flat potentials (see e.g. \cite{BasteroGil:2009ec}). This may be particularly relevant for supergravity and string theory models \cite{warm_string, warm_brane}, where one typically finds $m_\phi\gtrsim H$, precluding a sufficiently long period of inflation. 

We are mainly interested in dissipative effects in the {\it adiabatic} regime, where the microscopic particle dynamics is much faster than the evolution of any macroscopic variable, which typically holds during slow-roll inflation. The calculation of dissipation coefficients in the adiabatic regime has been a long studied problem in quantum field theory, starting primarily with the seminal works in the 80's of Hosoya and Sakagami \cite{hosoya1} for the $\phi^4$ interaction (see also \cite{morikawa1}). This was followed by Morikawa \cite{morikawa2}, who used the Closed Time Path formalism and obtained an effective Langevin-like equation, including an explicit fluctuation-dissipation relation. Fluctuation-dissipation relations emerging from quantum field theory models have since been
examined by several other authors \cite{hu1,boya1,boya2,GR}. Initially these treatments considered weak dissipation, leading to an underdamped evolution, although it was subsequently shown in \cite{Berera:1998gx} that strong dissipation with overdamped trajectories could also be achieved. All these studies were performed in Minkowski spacetime and one of the first to consider dissipation in curved spacetime was Ringwald \cite{ringwald}, with other subsequent treatments such as \cite{boya3,Lawrie:1998ic,BRfrw}. Dissipation from quantum field theory models has in fact been considered for a variety of applications beyond warm inflation dynamics, such as phase transitions, heavy ion collisions and conventional reheating after inflation.

{}For the reasons described above, the interactions between the inflaton and the radiation must be sufficiently suppressed to maintain the flatness of the potential, and in particular the finite temperature of the radiation may induce large thermal corrections to the inflaton mass, with $m_\phi\sim T>H$. This in fact precludes successful realizations of warm inflation in the simplest models, with the inflaton coupled directly to light scalars or fermions as described above, as the additional friction cannot overcome the increase in the inflaton's mass \cite{Berera:1998gx, Yokoyama:1998ju}. This of course assumes these are relativistic degrees of freedom, with $T\gg m_{\chi,\psi}$, but, as previously discussed, fields which are directly coupled to the inflaton tend to acquire large masses during inflation.

This has motivated implementations of warm inflation scenarios in the low-temperature regime, with $T<m_{\chi,\psi}$ (see e.g. \cite{others} for other recent implementations of warm inflation). In this case the on-shell production of these degrees of freedom is Boltzmann-suppressed, so that most models in the literature have so far focused on virtual modes, which may still lead to significant dissipative effects depending on the field multiplicities. In fact, this scenario, proposed in \cite{Berera:2002sp} and suggestively known as the {\it two-stage mechanism}, has all the ingredients of hybrid inflation models, where it is the waterfall field and not the inflaton that interacts with and may decay into light degrees of freedom. Moreover, it has been shown that the inclusion of both bosonic and fermionic degrees of freedom in a supersymmetric theory may help controlling both radiative and thermal corrections to the inflaton potential, despite supersymmetry being broken by both the finite energy density and the finite temperature during warm inflation \cite{Hall:2004zr}. 

In this work, we extend earlier computations of the dissipation coefficient $\Upsilon$ in the low-temperature regime \cite{Moss:2006gt, BasteroGil:2010pb} for the case where the fields coupled to the inflaton have an arbitrary number of possible decay channels, including both on-shell and off-shell production. Although most computations have to be performed numerically, our goal is to provide accurate expressions for $\Upsilon$ that may be used in constructing models of warm inflation, and determine the associated constraints on the field masses, couplings and multiplicities.

Our work is organized as follows. In the next section we describe the interactions between the inflaton field, the heavy fields and the light degrees of freedom that make up the radiation bath. We then discuss the leading thermal corrections to the particle masses, which may be found in Section 3. In Section 4, we describe the computation of the dissipation coefficient in the low-temperature regime, for both on-shell and off-shell modes, and find analytical expressions that accurately describe the numerical data in each case. We summarize our main results and discuss their impact on warm inflation model-building in Section 5. Three appendices are also included, where we provide more detailed discussions of some of the results used in our computations.


\section{Supersymmetric model}

We consider a generic supersymmetric model with chiral superfields $\Phi$, $X$ and $Y_i$, $i=1,\ldots,N_Y$, described by the superpotential~\cite{BasteroGil:2009ec, Hall:2004zr}:
\begin{equation} \label{superpotential}
W={g\over2}\Phi X^2+{h_i\over2} XY_i^2+f(\Phi)~,
\end{equation}  
where a sum over the index $i$ is implicit.  The scalar component of the superfield $\Phi$ describes the inflaton field, with an expectation value $\phi=\varphi/\sqrt{2}$, which we assume to be real, and the generic holomorphic function $f(\Phi)$ describes the self-interactions in the inflaton sector. The structure of the superpotential is quite generic in renormalizable models, following from the simple assumption that not all degrees of freedom are directly coupled to the inflaton field, and may for example be naturally implemented in D-brane constructions and related gauge theories (see e.g. \cite{warm_brane} and references therein). The analysis that we will go through below concentrates mostly on the scalar and fermionic sectors, which are already sufficiently involved, but  we discuss the possibility of embedding this superpotential in a gauge theory in Section 5.

The Lagrangian density describing the interactions between the inflaton vev and the scalar components of the superfields $X$ and $Y_i$, denoted by $\chi$ and $\sigma_i$, respectively, is given by:
\begin{eqnarray} \label{scalar_lagrangian}
\mathcal{L}_{scalar}&=&V(\varphi)+{1\over2}g^2\varphi^2|\chi|^2+{g\over2}\sqrt{V(\varphi)}\left(\chi^2+\chi^{\dagger 2}\right)+{g^2\over4}|\chi|^4+\nonumber\\
&+&{h_i\over2}{g\varphi\over \sqrt{2}}\left(\chi\sigma_i^{\dagger 2}+\chi^\dagger\sigma_i^2\right)+{h_ih_j\over4}\sigma_i^2\sigma_j^{\dagger 2}+h_i^2|\chi|^2|\sigma_i|^2~,
\end{eqnarray}  
where $V(\varphi)=|f'(\phi)|^2$ is the potential driving inflation. Similarly, the interactions involving the fermionic components $\psi_\chi$ and $\psi_{\sigma_i}$ and the scalar inflaton are given by:
\begin{eqnarray} \label{fermion_lagrangian}
\mathcal{L}_{fermion}&=&{g\varphi\over\sqrt2}\bar\psi_\chi P_L\psi_\chi+h_i\chi\bar\psi_{\sigma_i}P_L\psi_{\sigma_i}+{h_i\over 2}\sigma_i\bar\psi_{\sigma_i}P_L\psi_\chi+\mathrm{h.c.}~,
\end{eqnarray}  
where $P_L=(1-\gamma_5)/2$ is the left-handed chiral projector. Note that for our study of the dissipative dynamics during inflation we are only interested in the interactions involving the inflaton vev, but one should take into account that there are also interactions involving scalar fluctuations about the background value and their fermionic superpartners, although, for simplicity, we do not write them explicitly in Eqs.~(\ref{scalar_lagrangian}) and (\ref{fermion_lagrangian}). Also, for the purposes of this work, we will assume that the couplings $g$ and $h_i$ are real, although complex couplings may play an important role in, for example, generating a baryon asymmetry through dissipative effects during warm inflation \cite{BasteroGil:2011cx}. For simplicity, we will also take $h=h_i$ for all light species in our discussion, although our results can be easily generalized to the case of distinct couplings.

{}From the interactions shown in Eqs.~(\ref{scalar_lagrangian}) and (\ref{fermion_lagrangian}), we can see that the fields in the $X$ multiplet, which are directly coupled to the inflaton in the superpotential, acquire tree-level masses that are generically large for large values of the inflaton vev. On the other hand, the fields in the $Y_i$ multiplets remain massless at tree-level, although we may in general consider an additional mass term for these fields as we discuss in the next section. The non-vanishing potential energy driving inflation breaks supersymmetry, which leads to a mass splitting in the scalar sector of the $X$ multiplet:
\begin{eqnarray} \label{X_masses}
m^2_{\chi_{R,I}}&=&{g^2\varphi^2\over2}\left(1\pm{\sqrt{V(\varphi)}\over g\varphi^2}\right)~,\nonumber\\
m^2_{\psi_{\chi}}&=&{g^2\varphi^2\over2}~,
\end{eqnarray}
where $\chi=(\chi_R+i\chi_I)/\sqrt2$. Due to the smallness of the inflaton self-interactions typically required for a sufficiently long period of slow-roll inflation and a small enough amplitude for the spectrum of primordial density perturbations, this splitting may in general be neglected during inflation. {}For example, for $V(\varphi)=\lambda^2\varphi^4$, the splitting factor is $1\pm \lambda/g$, with the primordial spectrum imposing $\lambda\sim 10^{-7}$, so that we generically expect $\lambda\ll g$ (see e.g. \cite{BasteroGil:2009ec}). Henceforth we will then consider a common mass $m_\chi=g\varphi/\sqrt{2}$ for all fields in the $X$ multiplet. If the inflaton vev is sufficiently large, this mass will be above the temperature of the radiation bath during warm inflation, in which case thermal contributions to the inflaton potential from $\chi$ and $\psi_\chi$ loops are Boltzmann-suppressed, $e^{-m_\chi/T}\ll1$, and do not destroy the flatness of the 
tree-level potential.

This mass splitting may nevertheless be important when computing radiative corrections to the inflaton potential, and the supersymmetric Coleman-Weinberg potential has been shown to yield \cite{Hall:2004zr}:
\begin{equation} \label{Coleman-Weinberg}
V_{CW}={1\over 32\pi^2}\mathrm{Str}\left[\mathcal{M}_X^4\left(\ln\left({\mathcal{M}_X^2\over\mu^2}\right)-{3\over2}\right)\right]\simeq {g^2\over32\pi^2}V(\varphi)\ln\left({m_\chi^2\over\mu^2}\right)~,
\end{equation}
where $\mu$ is the renormalization scale and we have taken the leading correction from supersymmetry breaking. This shows that the leading radiative corrections to the inflaton potential are logarithmic and hence will not spoil the flatness of the potential for $g\lesssim 1$. Note that if we consider $N_X$ multiplets coupled to the inflaton as in Eq.~(\ref{superpotential}), the radiative corrections will be proportional to $g^2N_X$, and this may still accommodate a moderately large field multiplicity without a significant tuning of the coupling constant.

One should also notice that in some cases the mass splitting induced by supersymmetry breaking may actually become relevant at the end of the inflationary evolution. {}For example, for $f(\Phi)=g M^2\Phi$, we have $V(\varphi)\simeq g^2M^4$ and $m^2_{\chi_{R,I}}=m_\chi^2\pm M^2/2$, so that the imaginary component becomes tachyonic for $\varphi<M$, leading to a phase transition that ends inflation in the conventional hybrid mechanism. 

The superpotential (\ref{superpotential}) is thus a simple generalization of supersymmetric hybrid inflation that allows for an arbitrary inflaton potential and additional couplings of the waterfall field(s) to other scalars and fermions. These couplings allow for the decays $\chi\rightarrow \sigma\sigma, \psi_\sigma\psi_\sigma$ and $\psi_\chi\rightarrow\sigma\psi_\sigma$, as can be easily seen from Eqs.~(\ref{scalar_lagrangian}) and (\ref{fermion_lagrangian}), and thus allow the waterfall field and its superpartners to `reheat' the universe after the hybrid transition. Most importantly, both real and virtual $\chi$ and $\psi_\chi$ pairs can be produced by the rolling inflaton field through non-local quantum effects, and in turn decay into the $Y_i$ multiplet particles, leading to dissipative particle production and to the two-stage realization of warm inflation \cite{Berera:2002sp}, which will be the main topic of this work.

The interactions in Eqs.~(\ref{scalar_lagrangian}) and  (\ref{fermion_lagrangian})  also lead to scattering processes that help keeping the light fields in a state close to thermal equilibrium. Although this is not a crucial assumption for the presence of dissipative effects, it allows one to compute the associated dissipative coefficient in terms of the temperature of the radiation and of the field masses and couplings. In appendix C, we list some of the processes responsible for thermalization of the $\sigma$, $\psi_\sigma$ and $\phi$ fields and estimate the magnitude of the corresponding thermalization rates. Our estimates show that the scalar $\sigma$ fields thermalize more quickly than their fermionic superpartners for $T\ll m_\chi$, which is a consequence of supersymmetry breaking during warm inflation. Furthermore, thermalization of the inflaton particle states takes longer due to the sequestering effect of the two-stage superpotential. However, as scattering rates generically grow with the temperature, $\Gamma_i\sim g_i^2 T$ for some effective coupling $g_i$, one expects that all particles may thermalize within a Hubble time for $T\gg H$, which we will assume in the remainder of our discussion.


\section{Finite temperature effects}

The two-stage interactions between the inflaton and $X$ and $Y_i$ fields lead to the formation of a thermal bath during inflation, and finite temperature effects may thus become important. Since $\chi$ and $\psi_\chi$ acquire large tree-level masses and we assume $T\ll m_\chi$, their contributions to thermal loop corrections are Boltzmann-suppressed and may be ignored. This precludes large thermal mass corrections not only to the inflaton field but also to the light $\sigma$ and $\psi_\sigma$ particles that form the radiation bath. 

\begin{figure}[htbp]
\centering\includegraphics[scale=0.6]{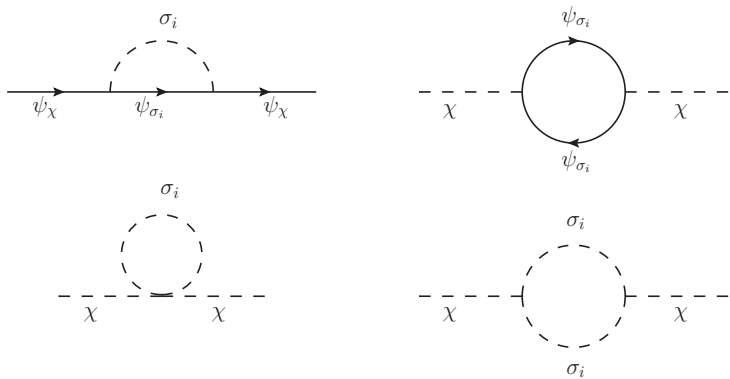} 
\caption{Feynman diagrams contributing to the self-energy of the $\psi_\chi$ and $\chi$ fields to leading order.}
\label{fig1}
\end{figure}

The masses of both the bosonic and fermionic components of the $X$ multiplet receive nevertheless corrections from their interactions with the light fields, the leading contributions corresponding to the first three diagrams in figure~\ref{fig1}.

These have been computed in \cite{Hall:2004zr} and are identical for both bosonic and fermionic degrees of freedom. Taking into account our normalization of the couplings in the superpotential (which differs from \cite{Hall:2004zr} by a factor $1/2$) and the field multiplicity, this yields the effective mass:
\begin{equation} \label{thermal_chi_mass}
\tilde{m}_{\chi}^{2}=m_\chi^2+{h^2N_Y\over 8}T^2~.
\end{equation}
In particular, this implies that the one-loop effective potential at finite temperature can be obtained by replacing the tree-level mass by $\tilde{m_\chi}$ in Eq.~(\ref{Coleman-Weinberg}). This shows that the leading thermal corrections to the inflaton potential are logarithmic and that, for $T\ll m_\chi$, they are smaller than the zero-temperature radiative corrections induced by the inflaton vev. We will nevertheless include these corrections in the computation of the dissipation coefficient. 

Although the last diagram in figure \ref{fig1}, involving a loop of the light $\sigma_i$ scalar fields, does not yield a $T^2$ correction, it yields a divergent contribution to the   $\chi$ mass that can be reabsorbed by a conventional zero-temperature renormalization procedure. The finite-$T$ contribution from this diagram was interpreted in \cite{Hall:2004zr} as a contribution to the temperature-dependent coupling $g(T)$. A closer inspection of this contribution reveals, however, a strong momentum dependence that goes beyond a simple coupling redefinition and modifies the two-point function for the $\chi$ scalars in a non-trivial way. As we discuss in detail in appendix A,  the full two-point function involves, in particular, a perturbative resummation of higher-order diagrams of this form, which is only valid in the regime $h^2N_Y\lesssim 1$. Hence, although thermal mass corrections would not be significant for a large number of decay channels $N_Y$, this correction places an upper bound on the effective coupling $h\sqrt{N_Y}$ that we must take into account in computing the associated dissipation coefficient.

\begin{figure}[htbp]
\centering\includegraphics[scale=0.6]{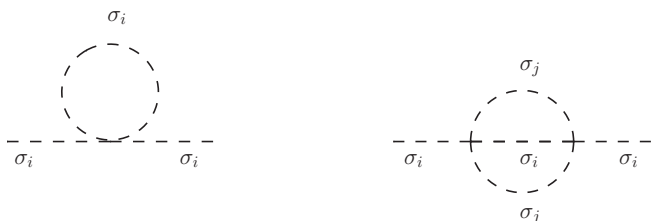} 
\caption{Feynman diagrams contributing to the self-energy of the $\sigma_i$ fields to leading order. Note that the `sunset' diagram on the right gives the leading-order correction involving different light species.}
\label{fig2}
\end{figure}

In the $Y_i$ sector, the light $\sigma_i$ scalars also receive thermal corrections from their self-interactions, while their superpartners only have couplings involving at least one heavy field and thermal loop contributions are suppressed in this case. The leading diagrams contributing to the scalar masses are shown in figure \ref{fig2}.

The leading corrections to the mass of each species $\sigma_i$ correspond to the quartic self-interactions $(h_i^2/4)|\sigma_i|^4$, while the mixing between different species, $(h_ih_j/4)\sigma_i^2\sigma_j^{\dagger 2}$ gives rise to the higher-order `sunset' diagram in figure \ref{fig2}, proportional to $h^4N_Y$ \footnote{Note that the relevant two-point correlation function for a complex scalar is $\langle\sigma_i\sigma_i^\dagger\rangle$, which receives distinct contributions from the quartic and bi-quadratic interaction terms.}. Since we will consider the regime $h^2N_Y\lesssim1$, which in turn requires $h<1$, it is sufficient to consider the leading corrections to the light scalar masses, given by \cite{Dolan:1973qd, Pinto:2006cb}:
\begin{equation} \label{thermal _sigma_mass}
m_\sigma^2=m_0^2+{h^2\over 12}T^2+\mathcal{O}(h^4N_Y)~,
\end{equation}
where we have included a possible tree-level mass arising from other potential SUSY breaking effects. If the light sector includes, for example, the MSSM fields, we expect $m_0\sim 1\ \mathrm{TeV}\ll T$ for the high temperatures typically involved in warm inflation models \cite{BasteroGil:2009ec}. Given that the leading thermal correction is independent of the field multiplicity, we thus expect $m_\sigma\ll T$, which we will assume in our discussion below. Furthermore, note that fields can be relativistic during warm inflation even if their mass is of the order of the Hubble parameter, since $T>H$, and in particular the inflaton particle states can be a part of the radiation bath if their thermal production is efficient (see appendix B).


\section{Dissipation coefficient}

The coupling between the inflaton and the $\chi,\ \psi_\chi$ fields induces time non-local corrections to the inflaton effective action which, in the adiabatic regime, where $\dot\varphi/\varphi, H\ll \Gamma_\chi$, with the latter denoting the $\chi$ decay width, lead to the effective friction term $\Upsilon \dot\varphi$ in the inflaton's equation of motion (see e.g.~\cite{BasteroGil:2010pb}). As mentioned above, these dissipative effects are due to the finite decay width of the $X$ multiplet fields, and correspond to an effective production of relativistic degrees of freedom, yielding a non-zero imaginary part for the self-energy of the fields in the $Y_i$ sector \cite{Graham:2008vu}.

\begin{figure}[htbp]
\centering\includegraphics[scale=0.7]{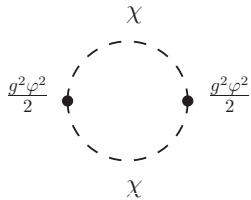} 
\caption{Feynman diagram contributing to the inflaton effective action at leading order.}
\label{fig3}
\end{figure}

The leading contribution to the dissipation coefficient from the complex scalar $\chi$ modes arises at one-loop order, as illustrated in figure \ref{fig3}, and has the following form \cite{BasteroGil:2010pb}:
\begin{equation} \label{dissipation_general}
\Upsilon={4\over T}\left({g^2\over 2}\right)^2\varphi^2\int {d^4p \over (2\pi)^4}\rho_\chi^2n_B(1+n_B)~, 
\end{equation}
where $n_B(p_0)=[e^{p_0/T}-1]^{-1}$ is the Bose-Einstein distribution and $\rho_\chi$ is the spectral function for the $\chi$ field. One expects the use of thermal distributions to be a good approximation to the dynamics of scalar fields coupled to a thermal bath in the adiabatic regime $\Gamma_\chi\gg H$, as discussed in \cite{Weldon:1983jn, Anisimov:2008dz, KT} (see appendix C).

The spectral function for the $\chi$ field entering in Eq. (\ref{dissipation_general}) corresponds to the fully dressed propagator, including the effect of their finite decay width into $\sigma_i$ particles:
\begin{equation} \label{spectral_function}
\rho_\chi(p_0,p)={4\omega_p\Gamma_\chi\over (p_0^2-\omega_p^2)^2+4\omega_p^2\Gamma_\chi^2}~, 
\end{equation}
where $\omega_p=\sqrt{\tilde{m}_\chi^2+p^2}$ for modes of 3-momentum $|\mathbf{p}|=p$ and energy $p_0$. Here we neglect the contributions of the real part of the $\chi$ self-energy and similar radiative corrections to the $|\phi|^2|\chi|^2$ vertex, which as discussed earlier and detailed in appendix A should hold for $h^2N_Y\lesssim 1$. 

The contribution from the $\psi_\chi$ modes is given by an analogous expression to Eq.~(\ref{dissipation_general}) with fermionic propagators and the relevant couplings, but has been shown to be suppressed in the low-temperature regime, $T\ll m_\chi$ \cite{BasteroGil:2010pb}, which is the main scope of this work. Thus, for the remainder of this work we will focus on the contribution from the $\chi$ field to the dissipation coefficient. It should be noted that the dissipation coefficient receives positive contributions from both the bosonic and fermionic fields that are directly coupled to the inflaton, so that only the time-local radiative corrections can be (partially) canceled by the underlying supersymmetry of the model. The suppression of the fermionic contribution is also a symptom of supersymmetry breaking during warm inflation, due to both the finite inflaton energy density and the finite temperature of the radiation bath.

The decay width of the $\chi$ bosons also includes contributions from both the bosonic and fermionic final states in the $Y_i$ multiplets. As we discuss below, the dissipative coefficient receives  contributions from both low-momentum and on-shell modes, and while in the latter case the bosonic and fermionic branching ratios are identical, for the former the fermionic decays are negligible (see e.g. \cite{BasteroGil:2010pb}). To simplify the computation of the dissipation coefficient, we will then focus on the bosonic channels, while nevertheless commenting on the effects of the fermionic $\psi_\sigma$ modes where appropriate.

The leading process contributing to the decay width of the $\chi$ fields is then the two-body decay $\chi\rightarrow \sigma_i\sigma_i$ (and the complex conjugate process), and at finite temperature we include contributions from both decays and inverse decays, as well as thermal scatterings off particles in the radiation bath. This has been computed in \cite{BasteroGil:2010pb} from the imaginary part of the $\chi$ self energy at one-loop order, yielding:
\begin{equation} \label{decay_width}
\Gamma_\chi={h^2N_Y\over 64\pi}{m_\chi^2\over \omega_p}F_T(p,p_0)~, 
\end{equation}
where 
\begin{eqnarray} \label{decay_width_function}
F_T(p,p_0)&=&\left[{\omega_+-\omega_-\over p}+{T\over p}\ln\left({1-e^{-{\omega_+\over T}}\over 1-e^{-{\omega_-\over T}}}{1-e^{-{p_0-\omega_-\over T}}\over 1-e^{-{p_0-\omega_+\over T}}}\right)\right]\theta\left(p_0^2-p^2-4m_\sigma^2\right)+\nonumber\\
&+&\left[{T\over p}\ln\left({1-e^{-{\omega_+\over T}}\over 1-e^{-{\omega_-\over T}}}{1-e^{-{p_0+\omega_-\over T}}\over 1-e^{-{p_0+\omega_+\over T}}}\right)\right]\theta\left(-p_0^2+p^2\right)~, 
\end{eqnarray}
with $\theta(x)$ denoting the Heaviside function and
\begin{equation} \label{omega_plusminus}
\omega_\pm=\sqrt{k_\pm^2+m_\sigma^2}~,\qquad k_\pm={1\over2}\left|p\pm p_0\left(1-{4m_\sigma^2\over p_0^2-p^2}\right)^{1\over2}\right|~. 
\end{equation}
Note that the factor $m_\chi^2$ appearing in Eq.~(\ref{decay_width}) arises from the fact that the coupling between the $\chi$ and $\sigma_i$ fields depends on the inflaton vev, whereas for the physical mass of the heavy bosons we use the effective mass $\tilde{m}_\chi$. Also, the first term in Eq.~(\ref{decay_width_function}) corresponds to direct and inverse decays, while the second term, corresponding to Landau damping, is associated to thermal scatterings.

\begin{figure}[htbp]
\centering\includegraphics[scale=0.9]{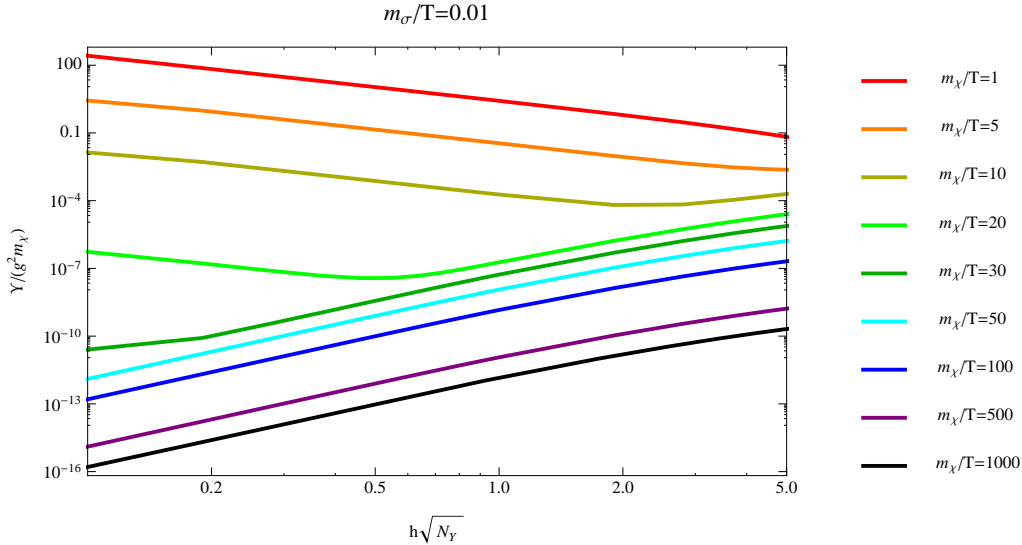} 
\caption{Numerical results for the dissipation coefficient as a function of the effective coupling $h\sqrt{N_Y}$ for different values of $m_\chi/T$. These results correspond to $m_\sigma/T=0.01$.}
\label{fig4}
\end{figure}

{}From Eqs.~(\ref{dissipation_general}) - (\ref{decay_width}), it is easy to see that the dissipation coefficient depends on the mass of $\chi$ fields, $m_\chi/T$, and the effective coupling, $h\sqrt{N_Y}$, up to an overall rescaling by the coupling $g^2$. We have integrated Eq.~(\ref{dissipation_general}) numerically for different values of these parameters, and we show a representative selection of our results in figure \ref{fig4}.

As one can easily conclude from the results shown in figure \ref{fig4}, the dependence on the effective coupling $h\sqrt{N_Y}$ is non-trivial, exhibiting two distinct behaviors for small and large couplings depending on the mass of the $\chi$ fields. As we will discuss below, these correspond to two different contributions to the integral in Eq.~(\ref{dissipation_general}) that can be analyzed separately and, in fact, represent different physical processes.


\subsection{Low-momentum contribution}

For larger values of the mass and effective coupling, the main contribution to the dissipation coefficient comes from {\it virtual} $\chi$ modes with low momentum and energy, $p,p_0\ll m_\chi$, so that one can use the approximation $(p_0^2-\omega_p^2)^2\simeq \tilde{m}_\chi^4$. If, in addition, these modes have a narrow width and thermal mass corrections can be neglected, $\Gamma_\chi\ll \tilde{m}_\chi\sim m_\chi$, the spectral function takes the simple form $\rho_\chi\simeq 4\Gamma_\chi/m_\chi^3$ and it is easy to see that:
\begin{equation} \label{dissipation_low_NWA}
{\Upsilon^{LM}\over g^2m_\chi}=Ah^4N_Y^2\left({T\over m_\chi}\right)^3~. 
\end{equation}
The constant can be determined from the numerical data and one obtains $A\simeq1.63\times10^{-3}$ for $m_\sigma/T=0.01$, consistently with the results obtained in \cite{Moss:2006gt, BasteroGil:2010pb}, taking into account our choice of normalization in Eq.~(\ref{superpotential}). 

\begin{figure}[htbp]
\centering\includegraphics[scale=0.85]{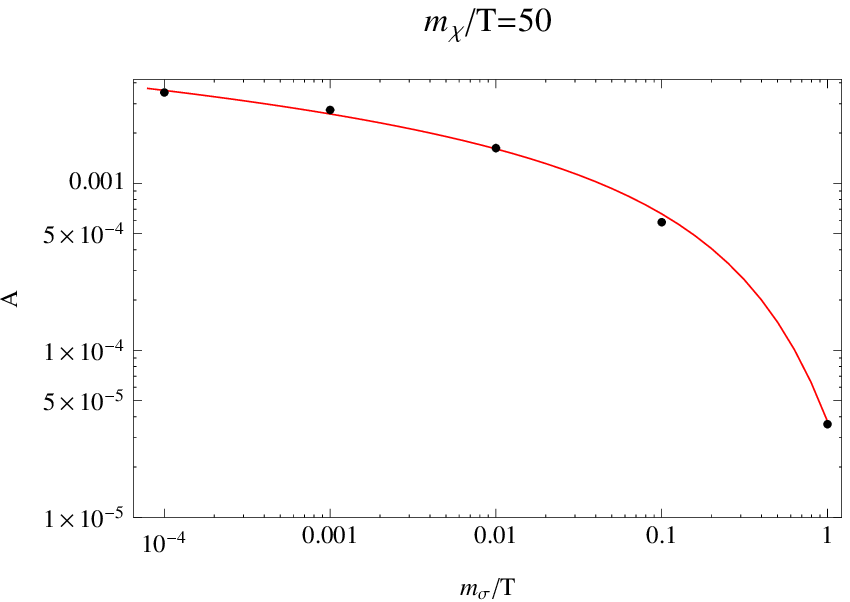} \hspace{0.5cm}
\centering\includegraphics[scale=0.85]{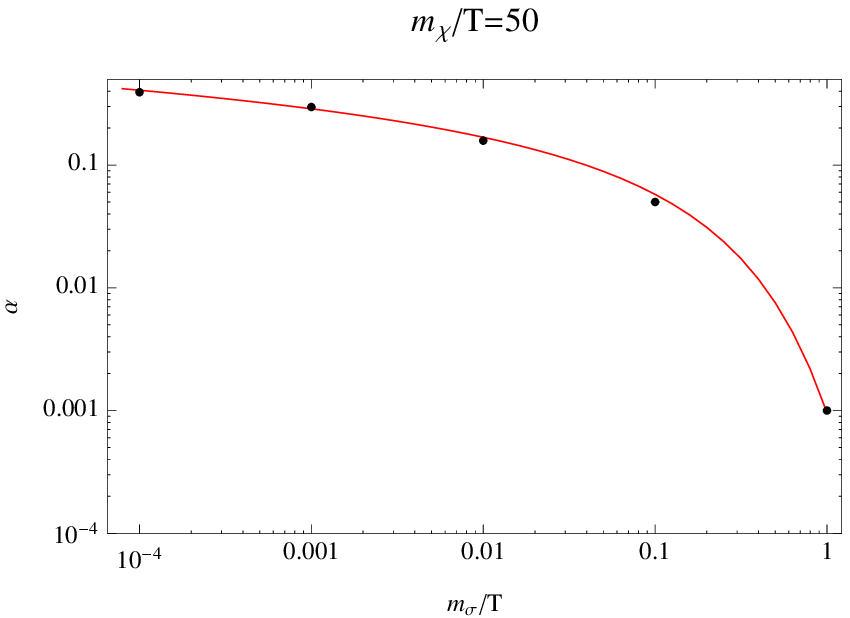} 
\caption{Dependence of the parameters $A$ and $\alpha$ on $m_\sigma/T$ and $m_\chi/T=50$. The solid lines represent the best fit curve in each case.}
\label{fig5}
\end{figure}

However, as one increases the effective coupling $h\sqrt{N_Y}$, the finite width of the $\chi$ fields and their thermally-induced mass become larger and modify the dissipation coefficient. Within the limit of our perturbative analysis, $h\sqrt{N_Y}\lesssim1$, this is well-described by a correction of the form: 
\begin{equation} \label{dissipation_low_corrected}
{\Upsilon^{LM}\over g^2m_\chi}={Ah^4N_Y^2\over1+\alpha h^2N_Y}\left({T\over m_\chi}\right)^3~, 
\end{equation}
with $\alpha\simeq 0.16$ for $m_\sigma/T=0.01$. The dependence of the $A$ and $\alpha$ parameters on the mass of the light fields is illustrated in figure \ref{fig5} for $m_\chi/T=50$.

The numerical data is then well-described by the following approximate expressions:
\begin{eqnarray} \label{parameters_sigma}
A&\simeq&-4.3\times10^{-4}\ln\left(1-e^{-{5\over2}{m_\sigma\over T}}\right)~, \nonumber\\
\alpha&\simeq&-0.05\ln\left(1-e^{-4{m_\sigma\over T}}\right)~.
\end{eqnarray}
This allows one to compute the $m_\sigma\ll T$ limit of the low-momentum contribution to the dissipation coefficient, which may be written as:
\begin{equation} \label{c_phi}
\Upsilon^{LM}\simeq C_\phi {T^3\over \varphi^2}~, \qquad C_\phi\simeq0.02h^2N_Y~. 
\end{equation}

Equation (\ref{c_phi}) is one of the main results of this work, showing that the dissipation constant grows like $C_\phi\propto h^2N_Y$ for light particles with
mass $m_\sigma \ll T$, as opposed to the $h^4N_Y^2$-dependence obtained in \cite{Moss:2006gt, BasteroGil:2010pb} for finite $m_\sigma/T$ and neglecting finite width effects. As discussed earlier, a perturbative analysis holds for $h\sqrt{N_Y}\lesssim 1$, so that this yields at most a dissipation constant $C_\phi\lesssim 1$ for a single $\chi$ scalar coupled to the inflaton field.


\subsection{Pole contribution}

The contribution to the dissipation coefficient from {\it real} $\chi$ modes is generically Boltzmann-suppressed, so that previous analyses assumed that it only becomes relevant in the high-temperature limit, $T\gg m_\chi$. However, our numerical results suggest that the Boltzmann suppression factor $e^{-m_\chi/T}$ may be compensated by a sufficiently small effective coupling in the low-temperature regime, so that on-shell modes become the dominant contribution to the dissipation coefficient, as can be seen in figure \ref{fig4}.

\begin{figure}[h] 
\centering\includegraphics[scale=1.1]{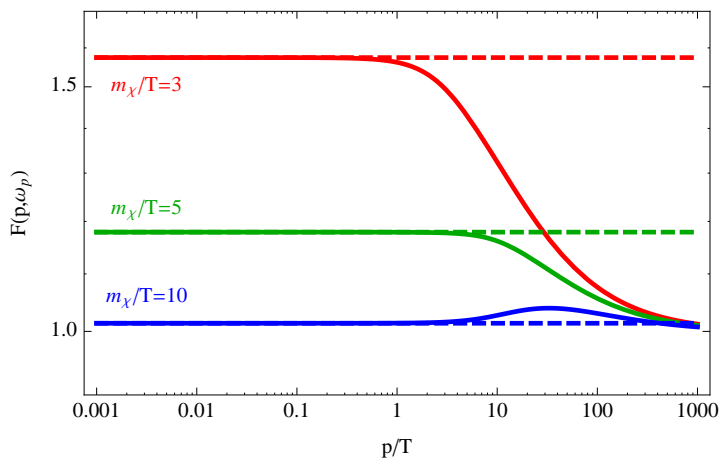}
\caption{The decay width function $F(p,\omega_p)$ for on-shell modes as a function of the 3-momentum for different values of $m_\chi/T\gtrsim1$. The solid lines yield the full function while the dashed lines correspond to the low-momentum approximation $F(p,\omega_p)\simeq 1+2e^{-{1\over2}{m_\chi\over T}}$.}
\label{fig6}
\end{figure}

To better understand this, let us expand the spectral function in Eq.~(\ref{spectral_function}) about its pole at $p_0\simeq \omega_p$, which corresponds to the production of on-shell modes. We then obtain:
\begin{equation} \label{dissipation_pole}
\Upsilon^P={2\over T}\left({g^2\over 2}\right)^2\varphi^2\int {d^3p \over (2\pi)^3}{1\over \Gamma_\chi \omega_p^2}n_B(1+n_B)~, 
\end{equation}
where $n_B\equiv n_B(\omega_p)$. For on-shell $\chi$ modes, decays into light scalars and fermions are equally probable, with partial decay widths given by Eq.~(\ref{decay_width}) for $p_0=\omega_p$. Although, for simplicity of the numerical calculation, we are focusing on the scalar decay modes, it is clear that in this regime the inclusion of the full decay width will lead to half the dissipation coefficient computed with only the scalar channels. Other decay channels may become relevant for small values of $h\sqrt{N_Y}$, but as discussed in appendix B we expect them to be generically sub-leading. Furthermore, for on-shell modes, the function in Eq.~(\ref{decay_width_function}) yields, at low-momentum $p$ and $m_\chi\gtrsim T$, in the limit $m_\sigma \ll T$:
\begin{equation} \label{F_function}
F(p,\omega_p)\simeq 1+2e^{-{1\over2}{m_\chi\over T}}+\mathcal{O}\left({p\over T}\right)^2~,
\end{equation}
which implies that thermal corrections to the decay width are subdominant in this limit. For $p\gtrsim T$ thermal corrections may become slightly larger, but as shown in figure \ref{fig6} they are generically sub-leading.

\begin{figure}[h]
\centering\includegraphics{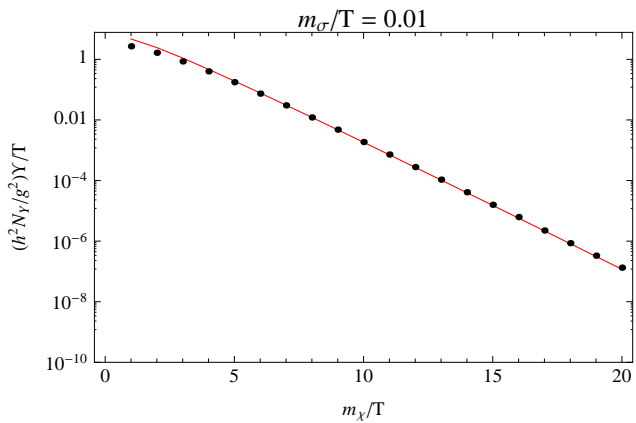}
\caption{Dissipation coefficient in the pole-dominated regime as a function of $m_\chi/T$ for $m_\sigma/T=0.01$. The points denote the full numerical result and the red line corresponds to the analytical expression in Eq.~(\ref{dissipation_pole_3}).}
\label{fig7}
\end{figure}

It is thus a good approximation to use the zero-temperature result for the decay width $\Gamma_\chi\simeq (h^2N_Y/64\pi)(m_\chi^2/ \omega_p)$ and the dissipation coefficient reduces to:
\begin{equation} \label{dissipation_pole_1}
\Upsilon_P\simeq{32\over \pi}{g^2\over h^2N_Y}\left({1\over T}\right)\int_0^\infty{p^2 dp\over \omega_p}n_B(1+n_B)~.
\end{equation}
Also, for $m_\chi\gg T$, we have, defining $x=p/T$:
\begin{eqnarray} \label{Bose_distribution}
n_B(1+n_B)\simeq& e^{-\sqrt{x^2+(m_\chi/T)^2}}\simeq e^{-m_\chi/T}e^{-{1\over2}{x^2\over m_\chi/T}}~,
\end{eqnarray}
where the second line follows from the fact that only low-momentum values give a significant contribution to the integral. Hence, we obtain:
\begin{equation} \label{dissipation_pole_2}
\Upsilon_P\simeq{32\over \pi}{g^2\over h^2N_Y}T e^{-m_\chi/T} I\left({m_\chi\over T}\right)~,
\end{equation}
where 
\begin{equation} \label{integral}
I(z)=\int_0^\infty dx{x^2\over\sqrt{ x^2+z^2}}e^{-{x^2\over 2z}}\simeq \sqrt{\pi z\over 2}~,
\end{equation}
where the last expression gives the leading result for $z\gg 1$. We thus obtain the following expression for the pole contribution to the dissipation coefficient:
\begin{equation} \label{dissipation_pole_3}
{\Upsilon_P\over T}\simeq{32\over \sqrt{2\pi}}{g^2\over h^2N_Y}\sqrt{m_\chi\over T} e^{-m_\chi/T} ~.
\end{equation}
Figure \ref{fig7} shows a comparison between the numerical results for the full dissipation coefficient in the pole-dominated region and the approximate expression derived above. One can see that Eq.~(\ref{dissipation_pole_3}) is in excellent agreement with the numerical results for $m_\chi/T\gtrsim 3$ and slightly overestimates the dissipation coefficient for smaller values. This is thus a very good approximation for practical purposes, since thermal corrections to the inflaton mass are not significantly suppressed for $m_\chi/T< 3$.

One should note that this contribution to the dissipation coefficient is more significant for small effective coupling, $h\sqrt{N_Y}$, in which case the $\chi$ scalar has a small width. Dissipation results in an effective friction coefficient only in the adiabatic limit, $\dot\varphi/\varphi,\,H\ll \Gamma_\chi$, so in constructing models of warm inflation in this pole-dominated regime one must make sure that this condition is nevertheless satisfied, which is of course a model-dependent issue beyond the scope of our generic analysis. Also, it has been shown that a full resummation of ladder diagrams contributing at the same order to the dissipation coefficient is required in this limit, although this does not generically change its order of magnitude \cite{Jeon}. Furthermore, these additional contributions to the dissipation coefficient can only make it bigger, since they can only contribute positively to the dissipation. Therefore, we can also see Eq. (\ref{dissipation_pole_3}) as a lower bound for the dissipation coefficient in the pole regime.


\subsection{Full dissipation coefficient}

\begin{figure}[h]
\centering\includegraphics[scale=1.2]{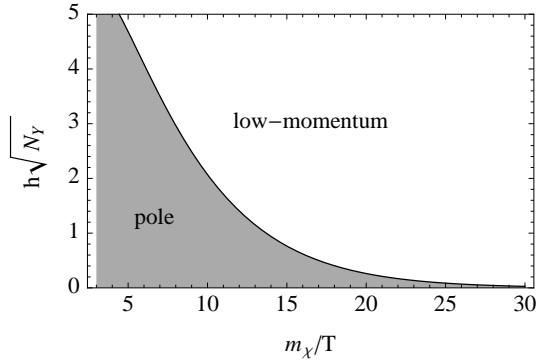}
\caption{Regions in the plane $(m_\chi/T, h\sqrt{N_Y})$ where the pole and the low-momentum contributions dominate the dissipation coefficient, with the solid line indicating the value of the effective coupling at which the contributions are comparable, given in Eq.~(\ref{crossover}).}
\label{fig8}
\end{figure}

The full dissipation coefficient receives contributions from both the low-momentum and near-pole regions in the $(p,p_0)$ plane of the intermediate $\chi$ fields and, putting together the results in Eqs.~(\ref{c_phi}) and (\ref{dissipation_pole_3}) we obtain:
\begin{eqnarray} \label{dissipation_total}
{\Upsilon\over g^2 T}\simeq {32\over \sqrt{2\pi}}{1\over h^2N_Y}\sqrt{m_\chi\over T} e^{-m_\chi/T}+0.01h^2N_Y\left({T\over m_\chi}\right)^2 ~,
\end{eqnarray}
which is valid in the limit $m_\sigma\ll T$ and for $h\sqrt{N_Y}\lesssim 1$. Note that for multiple fields in the $X$ sector, with similar couplings, the dissipation coefficient is multiplied by the corresponding factor $N_X$. The two contributions become comparable for
\begin{equation} \label{crossover}
h\sqrt{N_Y}\simeq 6 \left({m_\chi\over T}\right)^{5/8}e^{-{1\over 4}{m_\chi\over T}}~.
\end{equation}
In figure \ref{fig8} we illustrate the values of the regions in the parameter plane $(m_\chi/T, h\sqrt{N_Y})$ where each contribution dominates the dissipation coefficient, showing that for small (large) masses and couplings dissipation is dominated by real (virtual) modes.

\begin{figure}[htbp]
\centering\includegraphics[scale=1]{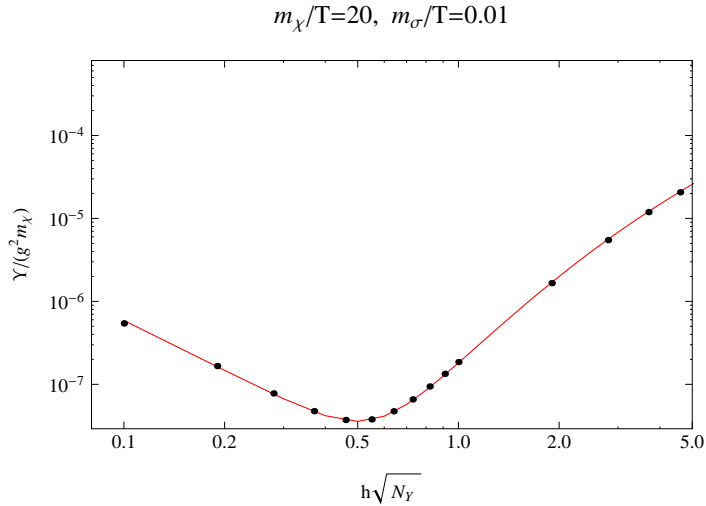} 
\caption{Comparison between the numerical results and the analytical expression obtained by adding the low-momentum and pole contributions to the dissipation coefficient for $m_\chi/T=20$ and $m_\sigma/T=0.01$.}
\label{fig9}
\end{figure}

In figure \ref{fig9} we compare the numerical results for the dissipation coefficient with the expression obtained by adding the pole and low-momentum contributions for $m_\sigma/T=0.01$ and $m_\chi/T=20$, illustrating the very good agreement between them.

\begin{figure}[htbp]
\centering\includegraphics[scale=1]{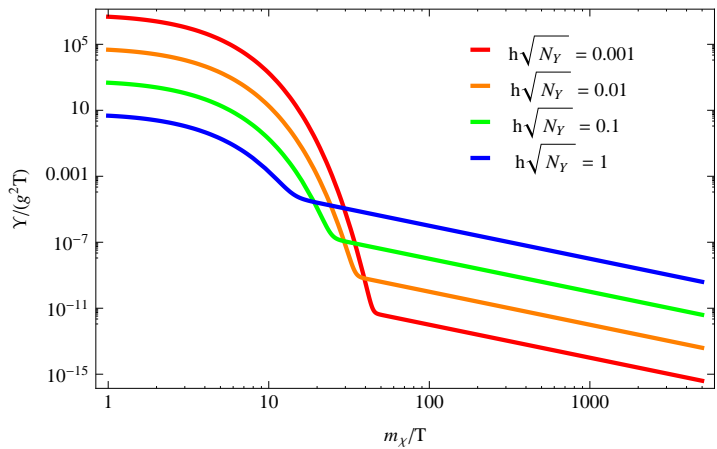} 
\caption{Full dissipation coefficient as a function of $m_\chi/T$ for different values of the effective coupling $h\sqrt{N_Y}$, using the expression inferred from the numerical results in Eq.~(\ref{dissipation_total}).}
\label{fig10}
\end{figure}

The most interesting result that follows from our analysis is the fact that the contribution of on-shell modes can be several orders of magnitude larger than the contribution of virtual modes to the dissipation coefficient, in particular given that this has so far been neglected in the low-temperature regime. This is of course due to the sharp $\chi$ resonance being able to compensate the Boltzmann suppression for any value of $m_\chi/T$ provided the effective coupling is sufficiently small. As shown in figure \ref{fig10}, one can easily obtain strong dissipative effects $\Upsilon\gg H$ for parameter regimes where $1\lesssim m_\chi/T\lesssim 10$ and $T>H$.

\begin{figure}[htbp]
\centering\includegraphics[scale=1]{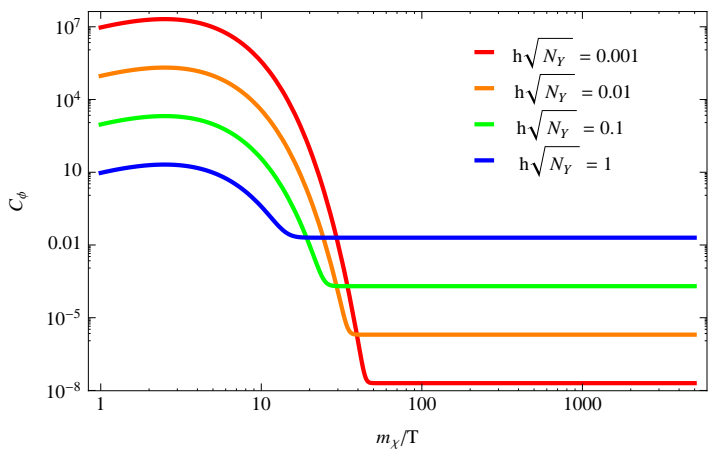} 
\caption{Effective dissipation parameter $C_\phi$ as a function of $m_\chi/T$ for different values of the effective coupling $h\sqrt{N_Y}$, using the expression inferred from the numerical results in Eq.~(\ref{dissipation_total}).}
\label{fig11}
\end{figure}

For a better comparison with results in the literature, we have computed an effective dissipation parameter $C_\phi$, illustrated in figure \ref{fig11},  by writing the full dissipation coefficient in the form $\Upsilon=C_\phi T^3/\varphi^2$.

In earlier analyses of the dynamics of warm inflation in the low-temperature/low-momentum regime (see e.g. \cite{BasteroGil:2009ec}), one generically requires $C_\phi\gtrsim 10^6$ in order to obtain a sufficiently long period of inflation with typical potentials. From our results, this would require a very large field multiplicity in the $X$ sector for virtual modes, whereas the on-shell contribution to the dissipation coefficient can easily be of this order for a single field and not too small effective couplings, as evident in figure \ref{fig11}. Of course the dissipation parameter $C_\phi$ is no longer a constant in this case, but this strongly motivates investigating in more detail the dynamics of warm inflation in this regime.


\section{Conclusion}

In this work we have extended earlier analyses of dissipative effects in supersymmetric warm inflation in the low-temperature regime,  $m_\chi\lesssim T$, to allow for an arbitrary number of decay channels for the intermediate fields. In agreement with earlier analyses, we find numerically that the dissipation coefficient $\Upsilon$ arising from the two-stage scalar interactions $\phi\rightarrow \chi\rightarrow \sigma_i$ differs significantly depending on whether the heavy fields $\chi$ are produced on- or off-shell, having determined analytical expressions for $\Upsilon$ in both cases.

Dissipation through virtual low-momentum $\chi$ modes is the dominant process for larger masses and couplings, leading to a dissipation coefficient of the form $\Upsilon=C_\phi T^3/\varphi^2$, with $C_\phi=0.02 h^2N_Y N_X$ if we allow for multiple fields in the $X$ sector coupled directly to the inflaton. This expression holds in the limit $m_\sigma\ll T$, which is valid for small Yukawa couplings $h$ even if the number of $Y_i$ species is large, and differs from the one obtained in \cite{BasteroGil:2010pb} for finite $m_\sigma/T$ due to finite width and mass corrections. Most importantly, our analysis is limited to the regime $h^2N_Y\lesssim 1$, as otherwise radiative corrections to the $\chi$ two-point function cannot be perturbatively resummed. This implies that a large number of light fields does not necessarily enhance the dissipation coefficient, since the system becomes effectively strongly coupled and a non-perturbative analysis is required. Since $C_\phi$ is independent of the coupling between the inflaton and the heavy mediators $g$, it may nevertheless be possible to consider models with a large number of $\chi$ fields, enhancing the dissipation coefficient while keeping $g^2N_X<1$ such that logarithmic corrections to the inflaton potential can be neglected.

On the other hand, for smaller masses and effective coupling, $h^2N_Y\ll 1$, dissipation is dominated by on-shell $\chi$ modes. For $m_\chi\gg T$, the dissipation coefficient is Boltzmann-suppressed, so that this regime has been largely overlooked in the literature. However, the $\chi$ resonance is very narrow in this regime, producing a strong peak in the momentum distribution that enhances the dissipation coefficient, with $\Upsilon\propto 1/(h^2N_Y)$ and given in Eq.~(\ref{dissipation_pole_3}) for $m_\sigma\ll T$. In fact, our results show that the contribution of on-shell modes can be several orders of magnitude larger than in the low-momentum case, which opens up a new possibility of constructing models of warm inflation with only a few fields coupled directly to the inflaton. One must ensure, however, that the system is evolving adiabatically in this regime, but since
\begin{equation} \label{adiabatic_condition}
{\Gamma_\chi\over H}\simeq {h^2N_Y\over 64\pi}\left({m_\chi\over T}\right)\left(T\over H\right)~, 
\end{equation}
we expect that very small couplings are nevertheless allowed in the low-temperature regime for $T\gg H$. We also note that the full dissipation coefficient is well-described by a sum of the contributions from real and virtual modes, which become comparable for the values of the effective coupling and mediator mass given in Eq.~(\ref{crossover}).

While our results provide the basic features of two-stage or catalyzed dissipation in the low-temperature regime, where thermal corrections to the inflaton potential are suppressed, they also suggest other interesting extensions of the basic model. Firstly, it would be interesting to develop techniques to compute the dissipation coefficient for a strongly coupled plasma, where our perturbative analysis fails but methods based for example on the AdS/CFT conjecture \cite{Maldacena:1997re} may be useful. As we have shown, this may be relevant for systems with a large number of relativistic degrees of freedom even if individual couplings are small, such as for example in Grand Unified Theories or multi-brane systems \cite{warm_brane}. In this case we expect, for example, many-body decays to become relevant, enhancing the decay width of the mediators and possibly the dissipation coefficient in the low-momentum regime, as well as potentially modifying the dynamics of warm inflation and the associated observables. 

Secondly, although for simplicity we have not included gauge fields in our analysis, we expect that in more realistic models both the mediators and the light degrees of freedom will carry gauge charges, in particular in realizations of warm inflation in the MSSM or its extensions. For minimally coupled $\chi$ fields, we do not expect direct decays into gauge fields to be significant, since the only two-body process $\chi \to \chi \gamma$, although possible at finite temperature, is suppressed by the large $\chi$ mass. Gauge interactions may nevertheless lead to higher-order decay channels and mediate additional scattering processes, thus enhancing the thermalization rates that keep particles close to thermal equilibrium. Finally, although this latter assumption considerably simplifies the analysis, it would be interesting to investigate the behavior of two-stage dissipative systems with non-thermal distributions, in particular how the relative contribution of on- and off-shell modes and the overall dissipative coefficient is modified in such conditions.

Our results have thus open new avenues of research in warm inflation and we hope that they motivate further research along these lines, both from the model-building and computational perspectives, as well as in exploring the effects of dissipative dynamics in other problems in modern cosmology.


\acknowledgments
J.~G.~R.~would like to thank Einan Gardi and Jennifer Smillie for useful discussions. M.~B.~G. is partially supported by MICINN (FIS2010-17395) and ``Junta de Andaluc\'ia'' (FQM101). A.~B.~and J.~G.~R.~are supported by STFC. R.~O.~R.~is partially supported by research grants from Conselho Nacional de Desenvolvimento Cient\'{\i}fico e Tecnol\'ogico (CNPq) and Funda\c{c}\~ao Carlos Chagas Filho de Amparo \`a Pesquisa do Estado do Rio de Janeiro (FAPERJ).


\appendix

\section{Radiative corrections to the two-point function}

The two-point function for the $\chi$ scalar field receives radiative corrections from its interactions with the fields in the light sectors, and the associated spectral function can be resummed to yield:
\begin{eqnarray} \label{rhochi}
\rho_{\chi} (p_0, {\bf p}) &=& \frac{i}{-p_0^2 + {\bf p}^2 + m_{\chi}^2
  +  \Sigma_{\chi}(p)} -  \frac{i}{-p_0^2 + {\bf p}^2  +
  m_{\chi}^2 + \Sigma_{\chi}^*(p)} \nonumber \\ &=& \frac{2 {\rm Im}\Sigma_{\chi}(p_0,{\bf p})}
{\left[-p_0^2  + m_{\chi}^2
  +  {\rm Re} \Sigma_{\chi}(p)\right]^2 + \left[ {\rm Im}\Sigma_{\chi}(p_0,{\bf p}) \right]^2 }~,
\end{eqnarray}
which reduces to the form in Eq.~(\ref{spectral_function}) if the real part of the $\chi$ self-energy can be neglected. Apart from thermal mass corrections, which we have included in our computation, the $\chi$ self-energy receives contributions from the $\sigma$-loops illustrated in the last diagram of figure \ref{fig1} (bottom right), which grow like $h^2N_Y$ and thus limit the light field multiplicities and couplings. This contribution is given, in Euclidean space, by:
\begin{equation} \label{Schi1}
\Sigma_\chi(P) = -2 N_Y \left(\frac{h}{2}m_\chi\right)^2 \int \frac{d^4 K}{(2 \pi)^4} \frac{1}{K^2 + m_\sigma^2}
\frac{1}{(K+P)^2 + m_\sigma^2}~.
\end{equation}
While the imaginary part of this diagram yields the scalar decay width in Eq.~(\ref{decay_width}), the real part has been computed at finite temperature in \cite{Weldon:1983jn, nishi}  and is given by:
\begin{eqnarray} \label{self_energy}
\mathrm{Re}[\Sigma_\chi](p_0,p)&=&-{h^2N_Y m_\chi^2\over32\pi^2}\left\{\left[{1\over\epsilon}-\gamma_E+\ln\left({4\pi\kappa^2\over m_\sigma^2}\right)+2\right.\right.-\nonumber\\
&-& C\ln\left(\left|{C+1\over C-1}\right|\right)\left[\theta(p_0^2-p^2-4m_\sigma^2)+\theta(-p_0^2+p^2)\right]-\nonumber\\
&-&2C\left.\arctan\left({1\over C}\right)\left[\theta(-p_0^2+p^2+4m_\sigma^2)+\theta(p_0^2-p^2)\right]\right]-\nonumber\\
&-&{1\over  p}\int_0^\infty dk\;k{n(E_1)\over E_1}\ln\left({kp+(p_0^2-p^2)/2-p_0^2E_1^2\over kp-(p_0^2-p^2)/2-p_0^2E_1^2}\right)^2\times\nonumber\\
&\times&\left[\theta(p_0^2-p^2-4m_\sigma^2)+\theta(-p_0^2+p^2)\right]\bigg\}~,
\end{eqnarray}
where $p=|\mathbf{p}|$, $C=(1-4m_\sigma^2/(p_0^2-p^2))^{1/2}$ and $E_1=\sqrt{k^2+m_\sigma^2}$. Note that the first three lines in this expression correspond to the $T=0$ divergent contribution, which depends on the renormalization scale $\kappa$, while the final integral yields the finite-temperature contribution. In the limit $m_\sigma\ll T$, this simplifies to yield:
\begin{eqnarray} \label{self_energy_zero_mass}
\mathrm{Re}[\Sigma_\chi](p_0,p)&=&-{h^2N_Y m_\chi^2\over32\pi^2}\left[{1\over\epsilon}-\gamma_E+2-{\pi\over 2}+\ln\left|{4\pi\kappa^2\over p_0^2-p^2}\right|-I_T(p_0,p)\right]~,
\end{eqnarray}
where the finite temperature contribution is given by the integral
\begin{equation} \label{finite_T_integral}
I_T(p_0,p)={1\over p}\int_0^\infty dk\;n(k)\ln\left({kp+(p_0^2-p^2)/2-p_0^2k^2\over kp-(p_0^2-p^2)/2-p_0^2k^2}\right)^2~.
\end{equation}

Note that the $T=0$ divergence is momentum-independent, and therefore we only need a mass renormalization counterterm, while the momentum dependence yields a finite wavefunction renormalization. The physical mass of the $\chi$ field is then given by 
\begin{equation} \label{renormalized_mass}
m_{\chi,R}^2(\kappa)=\tilde{m}_\chi^2+\mathrm{Re}[\Sigma_\chi]^{T=0}(m_\chi,0)~,
\end{equation}
where $\tilde{m}_\chi$ includes the $T^2$-corrections from the scalar tadpole and fermionic diagrams in figure \ref{fig1} and is given in Eq.~(\ref{thermal_chi_mass}). Note that the scalar tadpole also contributes to the $T=0$ divergence and should formally be included in the renormalization procedure, although this contribution is absent for $m_\sigma\ll T$. The dependence of the physical mass on the scale $\kappa$ leads to the usual running of the coupling $g$ according to the Renormalization Group equations.

The $T$-dependent contribution to the two-point function has been interpreted in \cite{Hall:2004zr} as contributing to the effective coupling $g(T)$ at finite temperature, having estimated a logarithmic temperature dependence from the integral in Eq.~(\ref{finite_T_integral}). Since in the low-momentum regime the dissipation coefficient is independent of the value of this coupling, they argued that this redefinition does not have a significant impact on the computation. However, a numerical inspection of this integral reveals a much stronger momentum dependence, in particular in the low-momentum regime, which as we have shown is the dominant contribution to dissipation for large effective coupling $h^2N_Y$.

In the low-momentum region, the integral in Eq.~(\ref{finite_T_integral}) is larger for  $p_0\lesssim p$ and well approximated by the expression:
\begin{eqnarray} \label{integral_results}
I_T(p_0,p)\simeq
-4\left(1+\pi{T\over p}\right)\cot^{-1}\left({3\over 2\pi}{p\over T}\right)~.
\end{eqnarray}
In this case the $T$-independent contribution is subleading for a broad range of values of the renormalization scale, so that we have for $p\lesssim T$:
\begin{eqnarray} \label{self_energy_low_approx}
{\mathrm{Re}[\Sigma_\chi]^{LM}\over m_\chi^2}\simeq -{h^2N_Y\over16}{T\over p}~.
\end{eqnarray}
Since the form of the spectral function in Eq.~(\ref{rhochi}) corresponds to a resummation of 1-loop contributions of this form, we can use Eq.~(\ref{self_energy_low_approx}) to estimate the regime of validity of perturbation theory. Although it has a strong momentum dependence, our numerical computation of the dissipation coefficient in this regime suggests that the largest contribution to the dissipation coefficient comes from momentum modes $p\lesssim T$, so that we estimate that perturbation theory breaks down for $h^2N_Y\lesssim 1$. For larger values of the effective coupling, the $\chi$ resonance also becomes broader  and we expect the Breit-Wigner form of the spectral function to be modified. Although the strong-momentum dependence of the self-energy at finite temperature precludes a more precise estimate of its contribution, we use this result as a guide to limit the regime of validity of our computation of the dissipation coefficient. In fact, when including the real part of the self-energy explicitly in the calculation, we observe deviations from the results presented in this work for $h^2N_Y\gtrsim 1$, with negligible changes for smaller values both in the low-momentum- and pole-dominated regimes.

\begin{figure}[htb] 
\centering\includegraphics[scale=0.6]{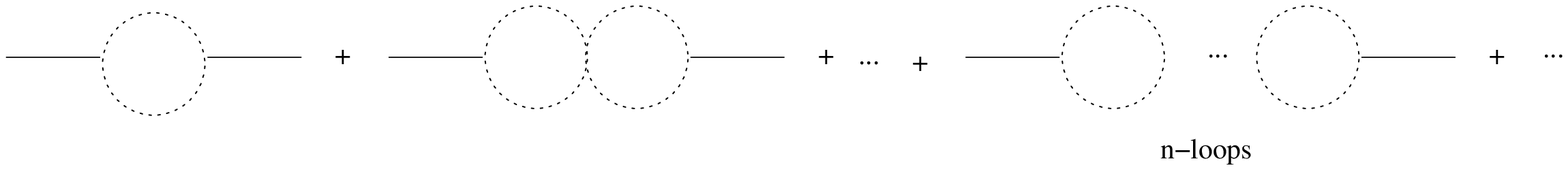}
\caption{Necklace-type $\sigma$ loops contributing to the $\chi$ self-energy.}
\label{fig12}
\end{figure}

One should also note that the self-energy receives  contributions from higher-order ``necklace" diagrams of the form illustrated in figure \ref{fig12}.

It is easy to see that a resummation of these higher-loop diagrams corresponds to a geometric series with ratio $\Sigma_\chi^{(1-loop)}/m_\chi^2$, so that our above estimate holds in this case as well. Furthermore, these diagrams will contribute to the $|\phi|^2|\chi|^2$ vertex, which is crucial for the dissipative dynamics of the inflaton field, emphasizing the importance of these corrections at large effective coupling.

Finally, it is worth mentioning that the fermionic loops in figure \ref{fig1} also contribute to $\mathrm{Re}[\Sigma_\chi]$ in a momentum-dependent way but, as the  fermionic couplings in Eq.~(\ref{fermion_lagrangian}) are dimensionless, their contribution is proportional to $T^2$ and hence suppressed in the low-temperature regime $T\ll m_\chi$.


\section{Decay channels}

The interactions in Eqs.~(\ref{scalar_lagrangian}) and (\ref{fermion_lagrangian}) yield multiple decay channels for the heavy $\chi$ scalars and which may contribute to dissipation. In general, we expect two-body decays into light scalars/fermions in the $Y$ sector to dominate the decay width $\Gamma_\chi$, but in the regime $h\sqrt{N_Y}\ll1$ we should explore the relative contributions of other channels. In this regime, the leading contribution to the dissipation coefficient corresponds to on-shell $\chi$ modes and finite-temperature corrections to the decay width are negligible, as discussed earlier. Furthermore, since $T\ll m_\chi$, it is sufficient to consider partial decay widths at zero temperature and momentum. The possible decay channels for on-shell $\chi$ bosons are then:

\begin{enumerate}
\item $\chi\rightarrow \sigma \sigma$
\item $\chi\rightarrow \psi_\sigma \psi_\sigma$
\item $\chi\rightarrow \sigma \sigma \phi$
\item $\chi_R\rightarrow \chi_I \phi_I$
\item $\chi_R\rightarrow \psi_\chi \psi_\phi$
\end{enumerate}

Notice that the first three decay channels involve both the real and imaginary parts of the complex $\chi$ scalars, while the last two are only possible due to the mass splitting in the $X$ sector from SUSY breaking during inflation and involve these components separately. In order to compute these contributions, it is useful to recall the general expression for scalar decays into pairs of scalars and fermions, $\chi\rightarrow \xi_1\xi_2$ and $\chi\rightarrow \psi_1\psi_2$, which at zero temperature and for $p_0=m_\chi$ and $p=0$ are given by \cite{BasteroGil:2010pb}:
\begin{eqnarray} \label{decay_generic}
\Gamma_{\chi}^{(S)}&=&{g_S^2\over 32\pi m_\chi}\sqrt{1-{(m_1-m_2)^2\over m_\chi^2}}\sqrt{1-{(m_1+m_2)^2\over m_\chi^2}}~,\nonumber\\
\Gamma_{\chi}^{(F)}&=&{g_F^2\over 8\pi}m_\chi\sqrt{1-{(m_1-m_2)^2\over m_\chi^2}}\left(1-{(m_1+m_2)^2\over m_\chi^2}\right)^{3/2}~,
\end{eqnarray}
where $g_{S,F}$ are generic couplings. The first two decay modes are then given by, for $m_\sigma\ll T$:
\begin{equation} \label{decay_light}
\Gamma_\chi^{(1,2)}={h^2N_Y\over 64\pi}m_\chi~.
\end{equation}
For the third decay channel, the three-body phase space yields a more complicated expression, but for the case where the inflaton mass can be neglected, which gives an upper bound to this contribution, we obtain:
\begin{equation} \label{decay_three_body}
\Gamma_\chi^{(3)}={h^2N_Yg^2\over 1024\pi^3}m_\chi~,
\end{equation}
which is naturally suppressed by the three-body phase space factor with respect to the first two decay channels for $g\lesssim 1$. 

The contribution from the channels (4) and (5) is somewhat model-dependent, depending on the SUSY mass splitting in the $X$ sector and hence the inflaton potential. The decay channel (4) corresponds to a term in the Lagrangian $g\chi_R\chi_I \mathrm{Im}[f(\phi)]$, so that the corresponding coupling is generically of the form $g_4=\lambda g \varphi$, where $\lambda$ is related to the inflaton self-coupling, up to numerical factors. On the other hand, the coupling determining channel (5) is independent of the inflaton potential and given by $g_5=g/(2\sqrt{2})$. If the inflaton/inflatino masses can be neglected compared to the mass splitting in the $X$ sector, this yields:
\begin{eqnarray} \label{decay_splitting}
\Gamma_\chi^{(4)}&=&{\lambda^2\Delta\over 16\pi} m_\chi~,\nonumber\\
\Gamma_\chi^{(5)}&=&{g^2\Delta^2\over 256\pi}m_\chi~,
\end{eqnarray}
where $\Delta=(m_{\chi_R}^2-m_{\chi_I}^2)/m_{\chi_R}^2$ and we used $m_{\chi_R}\simeq m_\chi=g\varphi/\sqrt{2}$. From Eq.~(\ref{X_masses}), we have $\Delta\simeq \lambda/g$ up to numerical factors, so that $\Gamma_\chi^{(4)}\propto \lambda^3$ and $\Gamma_\chi^{(5)}\propto \lambda^2$. Both decay channels are then suppressed with respect to (1,2) for $\lambda\ll h\sqrt{N_Y}$. 

For example, for a quartic potential $V(\varphi)=\lambda_{\varphi}^2\varphi^4$ we have $\lambda=\lambda_\varphi$ and $\Delta=2\lambda/g$, with observational constraints requiring $\lambda\simeq 10^{-7}$, as mentioned earlier. This implies that decays into the $Y$ sector will be dominant down to a very small Yukawa coupling. However, in this case we have $m_{\phi_I}=\lambda\varphi>m_{\chi_R}-m_{\chi_I}$ and $m_{\psi_\phi}=\sqrt{2}\lambda\phi>m_{\chi_R}-m_{\psi_\chi}$, so that both decays are actually kinematically forbidden. More generally, since the inflaton mass determines the slow-roll parameter $\eta_\phi=M_P^2V''/V$, these decays are forbidden for $\eta_\phi\gtrsim M_P/\varphi$, which is the case of monomial potentials. The relative contribution of these decay channels thus depends on the particular model of inflation considered, although in general we expect them to be subdominant due to the smallness of the SUSY mass splitting in the $X$ sector fields.

Finally, it should be mentioned that for off-shell $\chi$ bosons one finds additional decay channels at finite temperature, namely $\chi\rightarrow \chi\sigma\sigma$, $\chi\rightarrow \chi\phi\phi$ and $\chi\rightarrow \chi\chi\chi$, which are however Boltzmann-suppressed for $T\ll m_\chi$.


\section{Thermalization rates}

\begin{figure}[h]
\vspace{-1cm}
\includegraphics[scale=0.55]{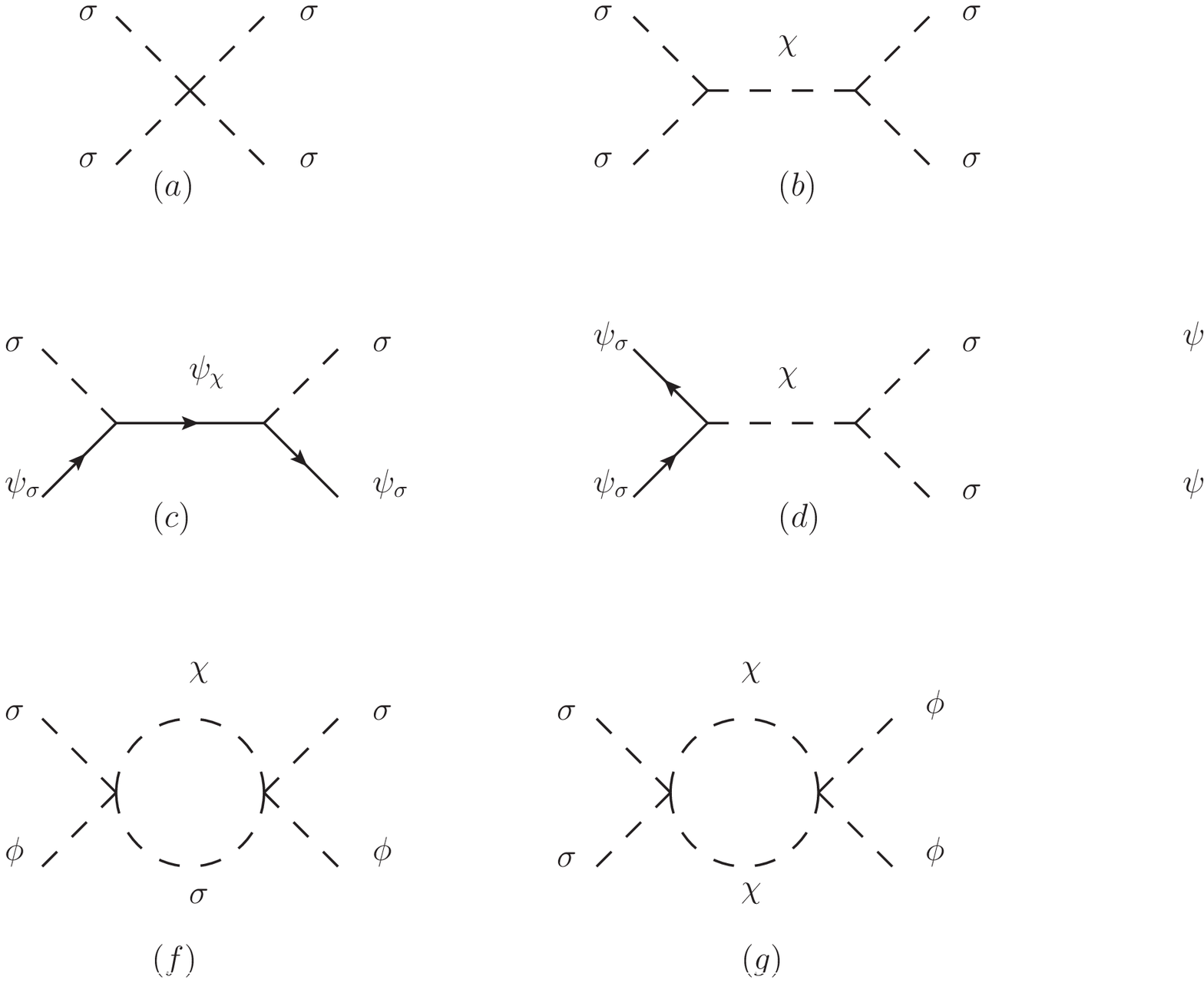}
\vspace{-6cm}
\caption{Feynman diagrams for scattering processes involving the light fields in the $Y$ sector and the scalar inflaton.}
\label{fig13}
\end{figure}

In computing the dissipation coefficient we have assumed that the relevant bosonic and fermionic fields are in a nearly-thermalized state. While in a non-expanding universe we expect interacting fields to eventually reach thermal equilibrium given enough time, in an expanding universe this requires thermalization processes to be faster than Hubble expansion. Although a comprehensive analysis of thermal scattering rates is beyond the scope of our discussion, here we list and estimate the rates for some of the main processes leading to thermalization of the light fields and also the inflaton particle states, assuming $T\gg m_\phi$. These are illustrated in figure \ref{fig13}.

For an arbitrary $2\rightarrow2$ process involving massless particles, the differential cross-section is given by \cite{Peskin}:
\begin{equation} \label{differential_cross_section}
{d\sigma\over d\Omega}={|\mathcal{M}|^2\over 64\pi^2 s}~,
\end{equation}
where $\sqrt{s}$ is the centre-of-mass energy. To estimate the relative magnitude of the different processes in figure \ref{fig13}, it is sufficient to neglect angular correlations and take typical momenta and energies, both for external and internal legs, to be of the order of the temperature of the thermal bath, $T\ll m_\chi$. This yields
\begin{equation} \label{thermal_cross_section}
\sigma_i\sim{|\mathcal{M}_i|^2\over 16\pi T^2}~,
\end{equation}
where $\mathcal{M}_i$ is the amplitude of the process. The rate at which a given species is thermalized is then given by $\Gamma_i=\langle\sigma_i n_i v\rangle$, and for relativistic particles we have $n_i=(f_i\zeta(3)/\pi^2)T^3$, with $f_i=2, 3/2$ for scalars and fermions, respectively, and $v\sim1$. Taking into account symmetry factors and initial state spin averages, we then obtain the following estimates for the processes in figure \ref{fig13}:
\begin{eqnarray} \label{cross_section_estimates}
\Gamma_{(a,b)}&\sim& {2\zeta(3)\over \pi^2}{h^4N_Y\over 32\pi}T~,\nonumber\\
\Gamma_{(c, d)} &\sim&{3\zeta(3)\over 2\pi^2}{h^4N_Y\over 32\pi}\left({T\over m_\chi}\right)^2T~,\qquad \Gamma_{(e)} \sim{3\zeta(3)\over 2\pi^2}{h^4N_Y\over 4\pi}\left({T\over m_\chi}\right)^4T \nonumber\\
\Gamma_{(f)} &\sim&{2\zeta(3)\over \pi^2}{h^4N_Y\over (4\pi)^416\pi}\left({T\over m_\chi}\right)^4T~,\qquad\!\! \!\!\!\! \!\!\Gamma_{(g)} \sim{2\zeta(3)\over \pi^2}{h^4N_Y\over (4\pi)^42\pi}\left({T\over m_\chi}\right)^8T 
\end{eqnarray}
This shows that the light $\sigma$ particles thermalize faster than their superpartners, which in turn thermalize more quickly than the scalar inflaton particles.  This was expected since the inflaton is sequestered from the light sector in the two-stage mechanism for $T\ll m_\chi$. Given that all reaction rates are of the form $\Gamma_i=g_i^2T$ all thermalization processes may be efficient for $T\gg H$ during warm inflation, even if the effective coupling $g_i$ is suppressed (see also \cite{Graham:2008vu} for a detailed discussion of the thermalization process). It may also be possible for additional light superfields $Z$ with no superpotential couplings to the $X$ sector, e.g.~$W_Z=h' YZ^2$, or for gauge interactions to enhance the thermalization rates of the fields in the $Y$ sector, which may be important in the small $h\sqrt{N_Y}$ regime, although a detailed analysis of this case is beyond the scope of the present work.

Note that the mediator fields are kept close to thermal equilibrium via the balance of decays, inverse decays and Landau damping processes for $\Gamma_\chi\gg H$. This can be seen explicitly in the Boltzmann equation for the $\chi$ field in the absence of dissipation, $\dot{n}_\chi+3Hn_\chi=-\Gamma_\chi(n_\chi-n_B)$ \cite{Weldon:1983jn, Anisimov:2008dz, KT}, which holds for all values of $m_\chi/T$ and for both off-shell and on-shell modes. Dissipation is, of course, an out-of-equilibrium process, but in the adiabatic regime where $\dot\varphi/\varphi,\ H<\Gamma_\chi$  entropy production is sufficiently slow for the system to remain close to thermal equilibrium at all times, thus justifying the use of  thermal distributions in the computation.


\end{document}